\documentclass[aps,prl,twocolumn,groupedaddress,showpacs]{revtex4}

\usepackage{graphicx}

\begin{document}
\title{Bose-Einstein condensates near a microfabricated surface}

\author{A.E. Leanhardt}
\author{Y. Shin}
\author{A.P. Chikkatur}
\author{D. Kielpinski}
\author{W. Ketterle}
\author{D.E. Pritchard}

\homepage{http://cua.mit.edu/ketterle_group/}

\affiliation{Department of Physics, MIT-Harvard Center for
Ultracold Atoms, and Research Laboratory of Electronics,
Massachusetts Institute of Technology, Cambridge, Massachusetts,
02139}

\date{\today}

\begin{abstract}

Magnetically and optically confined Bose-Einstein condensates were
studied near a microfabricated surface. Condensate fragmentation
observed in microfabricated magnetic traps was not observed in
optical dipole traps at the same location. The measured condensate
lifetime was $\geq 20$~s and independent of the atom-surface
separation under both magnetic and optical confinement.
Radio-frequency spin-flip transitions driven by technical noise
were directly observed for optically confined condensates and
could limit the condensate lifetime in microfabricated magnetic
traps.

\end{abstract}

\pacs{03.75.Fi, 03.75.Be, 39.20.+q, 34.50.Dy}

\maketitle

The manipulation of gaseous Bose-Einstein condensates with
magnetic fields produced by wires microfabricated on material
surfaces has opened a new frontier in the field of atom
optics~\cite{OFS01,HHH01,LCK02,SKH02}. Magnetic confinement using
microfabricated wires is tighter and has higher spatial resolution
than is achievable in macroscopic magnetic
traps~\cite{WEL95,SCH98,TOZ99,HIH99}.  Generally, an important
feature of magnetic traps is the excellent thermal isolation
between the nanokelvin temperature clouds they confine and the
300~K laboratory environment.  However, decreasing the
atom-surface separation into the micrometer range has raised
concerns that the surface may perturb the atoms through
fluctuating currents. Theoretical predictions suggest that
thermally induced atom-surface interactions will not pose
limitations for distances $\gtrsim 1\ \mu$m from the
surface~\cite{HEW99,HPW99}. While early demonstrations of trapping
and guiding of laser-cooled thermal atoms with microfabricated
devices reported no evidence of deleterious surface
effects~\cite{RHH99,MAG99,DLL00,FKC00}, recent studies using
samples cooled by forced radio-frequency (rf) evaporation to
$\lesssim 2\ \mu$K have found corrugated
potentials~\cite{LCK02,FOK02}, large heating
rates~\cite{HHH01,FOK02}, and short trap
lifetimes~\cite{HHH01,FOK02} for atom-surface separations in the
100~$\mu$m regime. The ultimate applicability of microfabricated
devices to atom optics depends on the characterization and
elimination of such effects.

In this work, we experimentally investigate the behavior of
Bose-Einstein condensates near a microfabricated surface.  The
condensates were confined at the same position relative to the
surface by either a microfabricated magnetic trap or an optical
dipole trap.  Since the two traps operate on different principles
and the electromagnetic fields for each have different sources,
this study provides a unique examination of the interaction
between Bose-Einstein condensates and a microfabricated surface.
For example, while condensates confined near the surface in a
microfabricated magnetic trap were found to fragment
longitudinally~\cite{LCK02,FOK02}, the clouds remained intact
under optical confinement.

Significantly, the measured condensate lifetime in both the
microfabricated magnetic trap and the optical dipole trap was
$\geq 20$~s, an order of magnitude longer than previous
results~\cite{HHH01,FOK02}, and independent of the atom-surface
separation. We have directly observed spin-flip transitions driven
by rf technical noise for condensates held in the optical dipole
trap.  The transition rate increased rapidly with decreasing
atom-surface separation implying that distance-dependent losses
can occur in magnetic traps where the products of such transitions
cannot be directly identified.

Bose-Einstein condensates containing over $10^7$ $^{23}$Na atoms
were created in the $|F=1,m_F=-1\rangle$ state in a macroscopic
Ioffe-Pritchard magnetic trap, loaded into the focus of an optical
tweezers beam, and transported $\approx 32$~cm in 2~s into an
auxiliary ``science'' chamber as described in Ref.~\cite{GCL02}.
The optical tweezers consisted of $\approx 50$~mW of 1064~nm laser
light focused to a $1/e^2$ radius of 26~$\mu$m. This resulted in
axial and radial trap frequencies $\omega_{||} = 2 \pi \times
4$~Hz and $\omega_\bot = 2 \pi \times 425$~Hz, respectively, and a
trap depth of 2.5~$\mu$K. Condensates containing $2-3 \times 10^6$
atoms arrived $70-500\ \mu$m below the microfabricated structures
mounted in the science chamber. The atom-surface separation was
varied by angling the optical tweezers axis before translation and
was limited to distances $\geq 70\ \mu$m due to the laser beam
clipping on the microchip support structures.

In the science chamber, the condensate either remained confined by
the optical tweezers or was loaded into a microfabricated
Ioffe-Pritchard magnetic trap formed by a Z-shaped wire carrying
current $I$ and an external magnetic bias field, $B_\bot$, as
described in Ref.~\cite{LCK02}.  An additional longitudinal bias
field, $B_{||}$, was applied with external coils to adjust the
magnetic trap bottom and radial trap frequency. The
microfabricated wires were lithographically patterned on a
600~$\mu$m thick silicon substrate mounted on an aluminum block.
They were 50~$\mu$m wide and electroplated with copper to a
thickness of 10~$\mu$m.

\begin{figure}
\begin{center}
\includegraphics{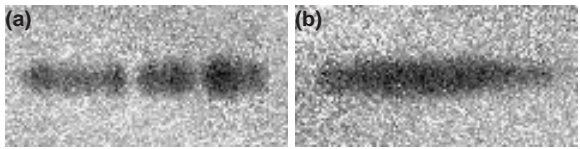}
\caption{Fragmentation of Bose-Einstein condensates. Radial
absorption images after 10~ms ballistic expansion of condensates
containing $\approx 10^6$ atoms after holding at a distance of
85~$\mu$m from the microfabricated surface for 15~s in the (a)
microfabricated magnetic trap and (b) optical dipole trap.
Longitudinal fragmentation occurred for condensates held in the
microfabricated magnetic trap, but not for those confined
optically at the same location with the microfabricated magnetic
trap off.  The microfabricated magnetic trap was operated with $I
= 130$~mA, $B_\bot = 3.2$~G, and $B_{||} = 1.4$~G yielding a
radial trap frequency $\omega_\bot = 2 \pi \times 450$~Hz.  The
optical dipole trap had a radial trap frequency $\omega_\bot = 2
\pi \times 425$~Hz and was operated with $B_{||}=1.8$~G. For both
condensates, the chemical potential was $\mu = k_B \times 120$~nK.
The absorption imaging light was resonant with the $F=1
\rightarrow F'=2$ transition. The field of view is 0.5~mm $\times$
1.0~mm.\label{f:fragments}}
\end{center}
\end{figure}

As in previous experiments~\cite{LCK02,FOK02}, condensates
confined near the surface in the microfabricated magnetic trap
were observed to fragment longitudinally
[Fig.~\ref{f:fragments}(a)].  The condensate density depletions
appeared in the same longitudinal position relative to the surface
on each realization of the experiment, and more fragments formed
as the atoms came closer to the microchip~\cite{LCK02}. In
contrast, condensates confined optically at the same location were
not observed to fragment [Fig.~\ref{f:fragments}(b)].  The same
longitudinal bias field was nominally applied to both magnetically
and optically confined condensates so that any surface
magnetization effects would perturb the clouds identically. The
lack of condensate fragmentation in the optical dipole trap
implies that the longitudinal potential corrugations arise due to
the presence of current flow in the microfabricated wires, in
agreement with conclusions reached elsewhere~\cite{KGO02}.
Deviations of the current flow from a straight line would lead to
such corrugations and could arise due to imperfect
microfabrication~\cite{LCK02} or current instabilities at high
current densities~\cite{KGO02}. The magnetic nature of the
potential corrugations and other possible origins are discussed in
Ref.~\cite{KGO02}.

It is interesting to note that in our earlier work no
fragmentation was observed when condensates confined in a
macroscopic Z-shaped wiretrap were brought within $\approx 10\
\mu$m of the surface of the wire~\cite{GCL02}.  The wire was made
of copper and had a circular cross-section with $1.27$~mm
diameter. The condensates were loaded into the wiretrap 740~$\mu$m
from the surface of the wire and brought closer by lowering the
wire current.  The experimental parameters upon closest approach
were $I=920$~mA and $B_\bot = 2.9$~G, yielding estimated axial and
radial trap frequencies $\omega_{||} = 2 \pi \times 7$~Hz and
$\omega_\bot = 2 \pi \times 78$~Hz, respectively~\cite{wiretrap}.
The macroscopic wiretrap contained $5 \times 10^5$ atoms extended
longitudinally over 200~$\mu$m at a chemical potential $\mu = k_B
\times 30$~nK. Differences between the macroscopic and
microfabricated wiretraps include vastly different fabrication
techniques as well as lower current densities in the macroscopic
wire.

Confined atoms are sensitive to noise at their trap frequency and
Zeeman splitting
frequency~\cite{HEW99,HPW99,SOT97,GOS98,JAU01,noise}.  In this
work, typical radial trap frequencies were $\approx 500$~Hz while
Zeeman splitting frequencies were $\approx 1$~MHz. Noise at the
trap frequency leads to heating and subsequent trap loss after the
atoms acquire an energy greater than the trap depth.  For atoms
confined in a Ioffe-Pritchard magnetic trap, radial magnetic bias
field fluctuations cause radial trap-center fluctuations.  The
amplitude of such trap-center fluctuations is independent of the
longitudinal bias field. However, for optically confined atoms,
only fluctuating radial magnetic field \emph{gradients} cause
radial trap-center fluctuations. The effects of such gradients can
be minimized by applying a longitudinal bias field that adds in
quadrature with the fluctuating radial gradients since it is the
gradient of the magnitude of the bias field vector that determines
the force on an atom.

Spin-flip transitions driven by rf noise at the atomic Zeeman
splitting frequency distribute the atomic population across
magnetically confinable and unconfinable states. This causes atom
loss for clouds held in magnetic traps. However, all spin states
are confined in an optical dipole trap so spin-flip transitions do
not lead to loss and the products can be directly observed. Since
magnetically and optically confined condensates react differently
to noise, whether it is at their trap frequency or the atomic
Zeeman splitting frequency, a systematic study of condensate
lifetimes in both magnetic and optical traps provides better noise
characterization than studies performed in either a magnetic or
optical trap exclusively.

\begin{figure}
\begin{center}
\includegraphics{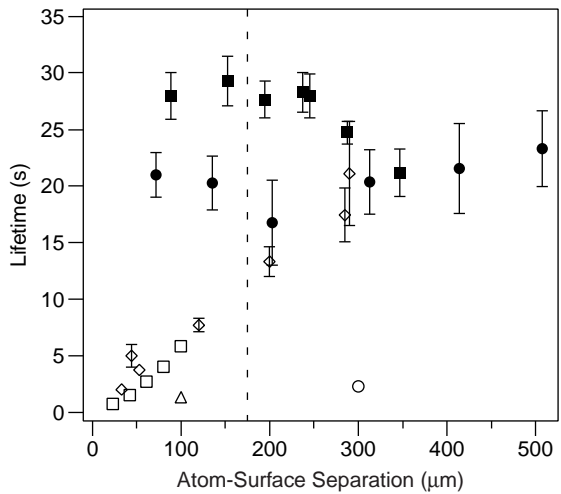}
\caption{Lifetime of Bose-Einstein condensates near a
microfabricated surface.  The $1/e$ lifetime of condensates
confined in the microfabricated magnetic trap (solid squares) and
optical dipole trap (solid circles) is shown to be independent of
distance from the microfabricated surface.  $I$ and $B_\bot$ were
varied with distance to maintain the radial magnetic trap
frequency at $\omega_\bot = 2\pi \times 450$~Hz with $B_{||} =
1.4$~G.  The vertical line indicates the onset of longitudinal
condensate fragmentation in the magnetic trap. In the optical
dipole trap, the condensate was held directly below the
microfabricated wire used to form the magnetic trap with $B_{||} =
1.8$~G.  No external connections were made to the microchip. The
optical dipole trap had axial and radial trap frequencies
$\omega_z = 2 \pi \times 4$~Hz and $\omega_\bot = 2 \pi \times
425$~Hz, respectively.  Only atoms remaining in the $|1,-1\rangle$
state were resonant with the absorption imaging light.  For
comparison, the distance dependence of thermal cloud lifetimes
measured in Ref.~\cite{FOK02} is shown for atoms confined
magnetically by a microstructure (open squares) and copper wire
(open diamonds). Error bars smaller than the symbol size are not
included. Also, magnetically confined condensate lifetimes
reported in Ref.~\cite{HHH01} (open triangle) and
Ref.~\cite{FOK02} (open circle) are shown for
comparison.\label{f:lifetime}}
\end{center}
\end{figure}

Any atom-surface coupling, regardless of frequency, should
manifest itself as a dependence of the condensate lifetime on the
atom-surface separation. Figure~\ref{f:lifetime} shows a
measurement of the magnetically and optically confined condensate
lifetime as a function of the distance from the microfabricated
surface.  No distance dependence was observed and the measured
condensate lifetime was $\geq 20$~s, ten times longer than
previous results~\cite{HHH01,FOK02}.  A distance independent
condensate lifetime indicates that atom-surface interactions are
unimportant over the $70-500\ \mu$m separation range~\cite{hinds}.
This is in contrast to results presented in Ref.~\cite{FOK02},
where a distance dependent lifetime was observed for thermal atoms
magnetically confined near a microfabricated surface. These data
are included in Fig.~\ref{f:lifetime} for comparison.

Several experimental details altered the measured condensate
lifetime. Excitations created during the microfabricated magnetic
trap loading were found to shorten the measured lifetime, and care
had to be taken to overlap the optical and magnetic traps during
transfer to minimize such excitations.  Translating the condensate
either towards or away from the microfabricated surface by
adiabatically varying $I$ and $B_\bot$ to shift the trap center
while maintaining a constant radial trap frequency was found to
decrease the condensate lifetime.  This presumably resulted from
excitations induced by irregular current changes due to technical
limitations in controlling the power supplies connected to the
microchip. As a result, microfabricated magnetic trap lifetime
data is only presented for atom-surface separations $\geq 70\
\mu$m, where the atoms were loaded into their final position
directly from the optical tweezers.  Occasionally, heating was
observed for atoms in both the microfabricated magnetic trap and
optical dipole trap due to technical noise at the trap frequency,
even with care taken to eliminate ground loops and minimize cable
lengths~\cite{grounding}. Connecting a 10~mF capacitor in parallel
with the 2~$\Omega$ microfabricated wire ($1/RC = 2 \pi \times
8$~Hz) eliminated such effects. Thereafter, applying rf power the
microchip at a frequency chosen to limit the trap depth for
magnetically confined atoms did not consistently alter the
condensate lifetime.

\begin{figure}
\begin{center}
\includegraphics{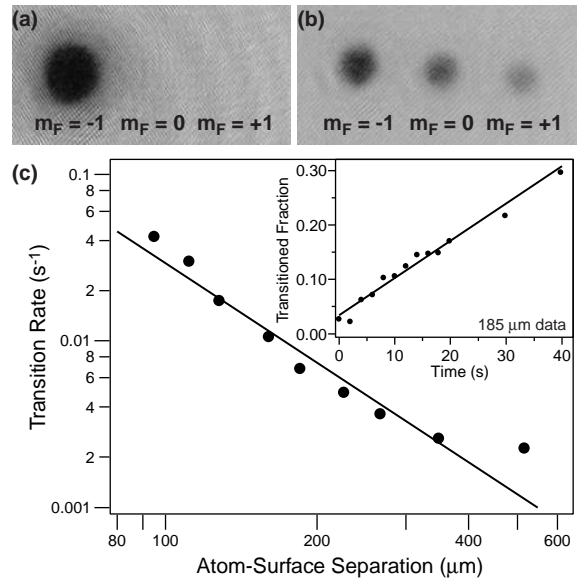}
\caption{Spin-flip transitions driven by radio-frequency technical
noise. Axial absorption images after 22~ms ballistic expansion of
condensates held in the optical dipole trap at a distance of
185~$\mu$m from the microfabricated surface for (a) 0~s and (b)
40~s.  A magnetic field gradient was applied during ballistic
expansion to separate the different spin states.  (c) Spin-flip
transition rate, $\Gamma$, vs distance, $d$, from the
microfabricated surface on a log-log scale.  A line $\Gamma
\propto 1/d^2$ is provided as a guide.  The inset shows the
fraction of the total atoms in the $m_F=0$ state with the
transition rate being defined as the initial slope of the data
(before any atoms in the $m_F=+1$ state were detected).  The
absorption imaging light was resonant with the $F=2 \rightarrow
F'=3$ transition. The atoms were optically pumped into the $F=2$
hyperfine level with a pulse resonant with the $F=1 \rightarrow
F'=2$ transition. This provided equal imaging sensitivity to each
magnetic sublevel. The field of view in (a) and (b) is 1.0~mm
$\times$ 2.0~mm.\label{f:spinflips}}
\end{center}
\end{figure}

The distance independent condensate lifetime presented in
Fig.~\ref{f:lifetime} indicates that our experiment is not
currently limited by the proximity of the microfabricated surface.
However, we have observed spin-flip transitions driven by rf noise
in the microfabricated wires. Figure~\ref{f:spinflips} shows the
behavior of condensates confined optically directly beneath the
microfabricated wire used for magnetic trapping. Condensate atoms
initially in the $|1,-1\rangle$ state [Fig.~\ref{f:spinflips}(a)]
were found to make transitions to other magnetic sublevels
[Fig.~\ref{f:spinflips}(b)].  Such transitions would act as a loss
mechanism for magnetically confined clouds.  The transition rate
was found to decrease as the square of the atom-surface separation
distance, $d$.  Since the magnetic field of a straight wire decays
as $1/d$, and the power scales as the square of the field, the
$1/d^2$ dependence of the spin-flip transition rate is expected
for atoms in the near field ($d \ll \lambda$) of the wire, where
$\lambda \approx 300$~m is the wavelength of $\approx 1$~MHz
radiation.

The transition rate vs distance data presented in
Fig.~\ref{f:spinflips}(c) was taken with all connections necessary
to run the microfabricated magnetic trap made to the microchip,
but with no current flowing in the microfabricated wires.  The
atoms were exposed to a longitudinal bias field $B_{||} =  1.8$~G
to nearly duplicate the field configuration in the microfabricated
magnetic trap. This also maximized their sensitivity to
fluctuating fields generated by wire currents since rf transitions
are more favorable for magnetic fields oscillating orthogonal to a
static bias field. Spin-flip transitions were suppressed by
exposing the optically confined atoms to an orthogonal bias field,
$B_\bot$.

The rf driven spin-flip transition rate depended strongly on
experimental details, suggesting that antenna effects coupled rf
noise into the system.  The rate was measured to be of order 100
times higher if care was not taken to carefully eliminate ground
loops and use minimal cable lengths~\cite{grounding}. Also, with
no connections to the microchip, spin-flip transitions were not
detectable for condensates held up to 60~s.  The spin-flip rate
presented in Fig.~\ref{f:spinflips}(c) became comparable to the
measured condensate decay rate displayed in Fig.~\ref{f:lifetime}
for atom-surface separations $< 100\ \mu$m.  Thus, extending long
condensate lifetimes much closer to the surface will require
further rf shielding and/or filtering.

In conclusion, we have studied the behavior of Bose-Einstein
condensates near a microfabricated surface. Condensates found to
fragment while held in a microfabricated magnetic trap were
observed to remain intact while held at the same position relative
to the microchip in an optical dipole trap. A possible explanation
is that deviations of the current path from a straight line give
rise to corrugations in the longitudinal potential. The origins of
such current path deviations are under investigation. Furthermore,
our work demonstrates magnetically and optically confined
condensate lifetimes $\geq 20$~s at distances $\geq 70\ \mu$m from
a microfabricated surface.  The lifetime was measured to be
independent of the atom-surface separation and ten times longer
than results obtained elsewhere at comparable distances. Spin-flip
transitions driven by rf technical noise were directly observed
for condensates held in an optical dipole trap, however we found
no evidence for fundamental, thermally induced noise driven
processes above the level of those attributed to technical noise.
Our results demonstrate the extreme sensitivity of Bose-Einstein
condensates to small static and dynamic electromagnetic fields.
This sensitivity provides a challenge for realizing
microfabricated atom-optical devices, but it also emphasizes the
potential for developing new detector and instrumentation
technology.

We thank T.~Pasquini for experimental assistance and
M.~Crescimanno for a critical reading of the manuscript. This work
was funded by ONR, NSF, ARO, NASA, and the David and Lucile
Packard Foundation. A.E.L.\ acknowledges additional support from
NSF.


\begin{thebibliography}{24}
\bibitem[{wir()}]{wiretrap}
\bibinfo{note}{The axial and radial trap frequencies were estimated by scaling
  the frequencies measured with $I = 2.0$~A in Ref.~\cite{GCL02} by
  (0.92~A/2.0~A)$^{1/2}$ and 2.0~A/0.92~A, respectively.}
\bibitem[{noi()}]{noise}
\bibinfo{note}{The atoms are sensitive to noise at multiples of their trap
  frequency depending on the symmetry of the excitation.}
\bibitem[{gro()}]{grounding}
\bibinfo{note}{The current supply for the microfabricated wire and the wire
  itself were grounded to the vacuum chamber at the location of the electrical
  feedthrough for the microchip connections. The total length of all
  (unshielded) current carrying cables was $\approx 1$~m.}
\end{thebibliography}
\end{document}